\documentclass[12pt,pra,aps,amssymb,amsfonts,amsmath,tightenlines]{revtex4}

\usepackage{dsfont}

\newcommand{\re}{\mbox{$\rm e$}}
\newcommand{\ri}{\mbox{$\rm i$}}
\newcommand{\rd}{\mbox{$\rm d$}}

\usepackage{xcolor}
\usepackage{bm}

\begin{document}

\title{Higher-order uncertainty bounds for mixed states}
\author{Alex J. Belfield$^{1}$ and Dorje C. Brody$^{1,2}$}
\affiliation{$^1$Department of Mathematics, University of Surrey, 
Guildford GU2 7XH, UK
\vspace{0.1cm}\\
$^2$St Petersburg National Research University of Information 
Technologies, Mechanics and Optics, St Petersburg 197101, Russia}

 %date commented out for arxiv submission
\date{\today}

\begin{abstract}
\noindent 
Uncertainty lower bounds for parameter estimations associated with a unitary family of mixed-state density matrices are obtained by embedding the space of density matrices in the Hilbert space of square-root density matrices. In the Hilbert-space setup the measure of uncertainty is given by the skew information of the second kind, while the uncertainty lower bound is given by the Wigner-Yanase skew information associated with the conjugate observable. Higher-order corrections to the uncertainty lower bound are determined by higher-order quantum skew moments; expressions for these moments are worked out in closed form. 
\end{abstract}

\maketitle

%%%%%%%%%%%%
%\large 

%%%%%%%%%%%%
\noindent \textbf{1. Introduction} \\

The Heisenberg uncertainty relation can be interpreted in a variety of ways 
\cite{KA,bush}, 
but perhaps operationally the most direct and mathematically the most 
straightforward way of understanding the relation is in the context of quantum 
state estimation. Take, for instance, the energy-time uncertainty relation 
$\Delta T \, \Delta H \geq \hbar/2$. The statistical interpretation of this relation 
is as follows. At time zero we prepare the system in a state, say, a pure state 
$|\psi_0\rangle$, and let the system evolve under the influence 
of the Hamiltonian ${\hat H}$. At a later point we wish to estimate how much 
time has elapsed since the initial preparation of the system. This amounts 
to estimating the parameter $t$ in the one-parameter family of states $|\psi_t 
\rangle=\re^{-{\rm i}{\hat H}t/\hbar}|\psi_0\rangle$ in Hilbert space. The 
estimation process is facilitated by use of an estimator ${\hat T}$, which is a 
maximally symmetric operator satisfying the properties that 
$\langle\psi_t|{\hat T}|\psi_t\rangle=t$ for all $t$ and that 
$\ri[{\hat H},{\hat T}]=\hbar$. The existence of such an operator ${\hat T}$, 
albeit not self-adjoint, for each given ${\hat H}$, is by now well 
understood \cite{holevo}. 

An analogous setup can be envisaged in the case of the position-momentum 
uncertainty relation $\Delta Q \, \Delta P \geq \hbar/2$. In this case the 
parametric family of states for the system, say, a particle, is given in the 
coordinate-space representation by 
$|\psi(q)\rangle=\re^{-{\rm i}{\hat P}q/\hbar}|\psi(0)\rangle$, with a given 
wave function $|\psi(0)\rangle$ at the origin of the parameter space. 
Assuming, without loss, 
that $\langle\psi(0)|{\hat Q}|\psi(0)\rangle=0$, then because the mean 
position $\langle\psi(q)|{\hat Q}|\psi(q)\rangle=q$ is given by the parameter 
$q$, the parameter estimation for the state $|\psi(q)\rangle$ amounts to 
estimating the position of the particle. 

In either case, to interpret the meaning of the uncertainty relation, we 
need to consider estimation errors. Specifically, in the context of parameter 
estimation in classical statistics, the quadratic 
error bound associated with an estimate is of great interest. In the classical 
context of parametric statistics one has a model, characterised by a parametric 
family of density functions $p(x|\theta)$ that is postulated to describe the 
data generated by sampling the value of a random variable $X$. The value 
of the parameter $\theta$, however, is unknown, and will have to be 
estimated from the given data set. If $\Theta(x)$ were an unbiased estimator 
for the parameter $\theta$ so that $\int \Theta(x) p(x|\theta)\,\rd x=\theta$, 
then by substituting the sampled value of $X$ in $\Theta(x)$ we get an 
estimate for $\theta$. 

To arrive at the lower bound for the quadratic estimation 
error (the variance), Rao \cite{Rao} considered the embedding of the density 
function in Hilbert space via the square-root map 
$\xi_\theta(x) = \sqrt{p(x|\theta)}$. Utilising the Dirac notation for 
(real) Hilbert space 
operations by writing, for instance, $|\xi_\theta\rangle=\xi_\theta(x)$ and 
$\langle \xi_\theta|{\hat\Theta}|\xi_\theta\rangle = \int \Theta(x) 
\xi_\theta(x)^2\,\rd x=\theta$, we have 
$\langle \xi_\theta|({\hat\Theta}-\theta) |\xi_\theta\rangle =0$. Differentiating 
this with respect to $\theta$, and using the relations $\langle{\dot\xi}_\theta 
|\xi_\theta\rangle=\langle\xi_\theta|{\dot\xi}_\theta \rangle=0$ and 
$\langle{\dot\xi}_\theta|{\hat\Theta}|\xi_\theta\rangle=
\langle\xi_\theta|{\hat\Theta}|{\dot\xi}_\theta \rangle$, we deduce that 
$\langle {\dot\xi}_\theta|({\hat\Theta}-\theta) |\xi_\theta\rangle =\frac{1}{2}$. 
Then from the Schwarz inequality $\langle{\dot\xi}_\theta|\eta\rangle^2 \leq 
\langle{\dot\xi}_\theta|{\dot\xi}_\theta\rangle\langle\eta|\eta\rangle$ we deduce 
by setting $|\eta\rangle=({\hat\Theta}-\theta) |\xi_\theta\rangle$ the 
Cram\'er-Rao inequality 
\begin{eqnarray}
\Delta\Theta^2 \geq \frac{1}{4\langle{\dot\xi}_\theta|{\dot\xi}_\theta\rangle} \, , 
\label{eq:1} 
\end{eqnarray} 
where we wrote $\Delta\Theta^2=\langle\xi_\theta|({\hat\Theta}-\theta)^2 
|\xi_\theta\rangle$ for the variance of the estimate. The quantity 
$4\langle{\dot\xi}|{\dot\xi}\rangle$ appearing on the right side is the Fisher 
information \cite{Fisher} (in the case of a multi-parameter family of densities 
the variance is replaced by the covariance matrix, and the Fisher information 
is replaced by the Fisher information matrix). The Cram\'er-Rao inequality 
(\ref{eq:1}) shows that the more information one can extract from sampling 
for the value of the unknown parameter $\theta$, the smaller the estimation 
error bound is. 

There is a second, geometric interpretation of the Fisher information, namely, 
that it gives the speed at which the state $|\xi_\theta\rangle$ changes in 
Hilbert space, as the parameter $\theta$ is varied. Thus, if the state 
$|\xi_\theta\rangle$ is sensitive to the parameter $\theta$ then the estimation 
error can be made small, whereas if the state hardly changes when the 
parameter is varied, then the estimation error will be large. 
The advantage of Rao's Hilbert space formulation of classical statistics is 
that it is readily applicable, \textit{mutatis mutandis}, to the problem of quantum 
state estimation. In particular, in the context of the above-posed state 
estimation problem, a short calculation making use of the 
Schr\"odinger-Kibble dynamical equation $|{\dot\xi}_t\rangle=-\ri\hbar^{-1}
({\hat H}-\langle{\hat H}\rangle)|\xi_t\rangle$ that removes the dynamical 
phase shows that $\langle{\dot\xi}_t|{\dot\xi}_t\rangle=\Delta H^2/\hbar^2$. 
It follows that the Cram\'er-Rao inequality 
reduces to the Heisenberg uncertainty relations \cite{BH1}. 

For quantum state estimation, starting from the Heisenberg relation there are 
two directions in which the analysis can be extended. The first concerns the 
investigation to obtain sharper bounds for the variance. For sure for some 
states the quadratic error is small (e.g., for a coherent state the position 
variance is equal to the inverse of the Fisher information), but for 
other states this is not the case. The higher-order corrections to the 
Heisenberg relation, leading to sharper uncertainty relations, were first 
obtained in \cite{BH1}, where by `order' we mean the degree of dispersion 
of the conjugate operator (but not in terms of, say, powers of Planck's 
constant so that contributions from higher-order terms need not be small). 
Thus, for instance, one can deduce that 
\begin{eqnarray}
\Delta Q^2 \, \Delta P^2 \geq \frac{\hbar^2}{4} \left( 1 + 
\frac{(\Delta P^4 - 3 (\Delta P^2)^2)^2}
{\Delta P^6 \, \Delta P^2-(\Delta P^4)^2} \right) 
\label{eq:2}
\end{eqnarray}
and so on \cite{BH2}, where we have written $\Delta P^n$ to mean the 
$n$-th central moment of ${\hat P}$. In fact, it is possible to work out an 
infinitely many such higher-order corrections explicitly \cite{BH3}. 

The second direction concerns the parameter estimation when the state of 
the system is described by a mixed-state density matrix. A mixed quantum 
state represents a probabilistic mixture of pure states. Thus, the problem of 
parameter estimation becomes more difficult, because on top of the intrinsic 
quantum-mechanical uncertainty represented by the pure states, there is an 
added, essentially classical, uncertainty regarding which pure state might be 
the `correct' state to describe the system. As a consequence, the Fisher 
information---the information that can be extracted from sampling, or 
measurements---is necessarily reduced from the variance of the conjugate 
variable, in the context of parameter estimation associated with a unitary 
curve in the space of mixed states. Specifically, it reduces to another information 
measure introduced by Wigner and Yanase \cite{WY}, as demonstrated in 
\cite{Luo,Luo2,DCB}. 

There is in fact a long history of research in developing a theoretical framework 
for investigating the quantum state estimation problem. In particular, starting 
with the work of Helstrom \cite{Helstrom}, the Fisher information associated 
with a parametric family of density matrices is often investigated by use of the 
technique of a symmetric logarithmic derivative. This is quite natural inasmuch 
as in classical statistics the Fisher information is commonly represented in 
terms of the log-likelihood function $\log p(x|\theta)$ (except in the work of Rao 
where it is represented in terms of the square-root density function 
$\sqrt{p(x|\theta)}$). Indeed, we have the identity 
\begin{eqnarray} 
\int p(x|\theta) \left( \frac{\rd\log p(x|\theta)}{\rd \theta}\right)^2 \rd x = 
4 \int \left( \frac{\rd\sqrt{p(x|\theta)}}{\rd \theta}\right)^2 \rd x \, .
\end{eqnarray}
Utilising the technique of symmetric logarithmic derivatives, Braunstein and 
Caves \cite{BC} maximised a representation for the Fisher information 
over all quantum states to show that for pure states the Cram\'er-Rao 
inequality reduces to the Heisenberg uncertainty relation. 

Although the use of symmetric logarithmic derivatives in the 
context of quantum state estimation has been very popular 
in the literature since the work in \cite{BC}, there are several limitations 
associated with this approach when attempting to obtain sharper uncertainty 
relations for mixed states. For the same token, the relation to the geometry of 
the underlying Hilbert space, as elegantly exploited by Rao in the case of 
classical statistics, becomes obscure. The point is that in classical parametric 
statistics one often works in the space of log-density functions, as opposed to 
the space of densities directly; the latter is just the subspace of the first 
Lebesgue class functions ${\cal L}^1({\mathds R})$. Thus, for instance, Efron's 
measure for information loss in higher-order asymptotic inference is given by 
the curvature of the parametric curve $l_\theta(x)=\log p(x|\theta)$ in the 
space of log-likelihood functions \cite{efron}. In contrast, Rao considered the 
Hilbert-space embedding of the parametric family of density functions, from 
which the more familiar machineries on the space ${\cal L}^2({\mathds R})$ 
of second Lebesgue class functions in mathematical analysis can be applied. 
In particular, the geometric meaning of the Fisher information then becomes 
immediately apparent \cite{Rao}. 

In a similar vein, when working with density matrices in quantum theory, one 
can either consider the log-density matrix or the square-root density matrix; 
but it is the latter that lends itself with the geometry of the associated Hilbert 
space. 
The idea is that by considering a Hermitian square root ${\hat\xi}$ of a 
density matrix ${\hat\rho}={\hat\xi}^2$ (any real root would suffice), we can 
embed the space of density matrices in a (real) Hilbert space of `pure' states 
${\hat\xi}$ equipped with the Hilbert-Schmidt (or just the trace) inner product. 
In contrast, the log-density matrix is not an element of Hilbert space. 
This line of thinking was exploited in \cite{DCB} to derive a modified form of 
uncertainty relations that arise naturally in the context of the Hilbert space of 
square-root density matrices. 

With these preliminaries, the purpose of the present paper is to derive 
higher-order corrections to the uncertainty lower bounds obtained in 
\cite{Luo,Luo2,DCB}. We shall follow, in particular, the approach developed 
in \cite{BH3}, but instead of pure states considered therein we will be 
working with square roots of mixed state density matrices. The analysis 
presented in what follows, however, are not mere generalisations of the 
work in \cite{BH3} from pure to mixed states in that, perhaps surprisingly, 
in the limit of pure states our results do not reproduce those of \cite{BH3}. 
Thus our results give rise also to a genuinely new set of higher-order 
corrections to the uncertainty lower bounds for pure states as well. 
More generally, for mixed 
states, we shall encounter modified forms of higher-order central moments 
that arise naturally. We shall refer to these 
as the higher-order `quantum skew moments' in analogy with the 
Wigner-Yanase skew information, which is a modified form of the second 
central moment (variance). We shall determine the form of these skew 
moments, and demonstrate how corrections to the 
uncertainty relation, to an arbitrary high order, can be derived explicitly in 
a recursive manner, expressed in terms of the various skew moments. \\ 

\noindent \textbf{2. Mixed state uncertainties and the skew information} \\

Let us begin by briefly reviewing the approach proposed in \cite{DCB} for 
benefits of readers less acquainted with the material. This 
will also be useful in setting the notations. 
Starting from a one-parameter family 
of states ${\hat\rho}_t=\exp(-\ri{\hat H}t) {\hat\rho}_0\exp(\ri{\hat H}t)$ 
associated with a prescribed initial state ${\hat\rho}_0$ we let ${\hat\xi}_t$ 
be an arbitrary Hermitian square-root of ${\hat\rho}_t$ so that 
${\hat\xi}_t^2={\hat\rho}_t$. The objective in state estimation here is to 
determine the elapse time $t$ since the initial preparation of the state. We 
let ${\hat T}$ be an unbiased estimator for the time parameter $t$ so that 
${\rm tr}({\hat T}{\hat\xi}_t^2)=t$. 

To proceed it will be useful to 
introduce the notation for the mean-adjusted operator by writing 
${\tilde T}={\hat T}-t{\mathds 1}$ so that the variance of ${\hat T}$ is given 
by $\Delta T^2= {\rm tr} ({\tilde T}^2{\hat\xi}_t^2)$. 
Differentiating the normalisation condition ${\rm tr}({\hat\xi}_t^2)=1$ 
in $t$ and writing ${\hat\xi}'_t= \partial_t {\hat\xi}_t$, we obtain 
${\rm tr}({\hat\xi}'_t {\hat\xi}_t)=0$. Differentiating 
${\rm tr}({\tilde T}\xi_t^2)=0$ in $t$ and using ${\tilde T}'=
-{\mathds 1}$ we find
${\rm tr} [({\tilde T}{\hat\xi}_t+{\hat\xi}_t{\tilde T}){\hat\xi}'_t]=1$. 
Applying the matrix Schwarz inequality, we deduce a form 
of the quantum Cram\'er-Rao inequality
\begin{eqnarray}
\Delta T^2 + \delta T^2 \geq \frac{1}{2\, 
{\rm tr}({\hat\xi}'_t{\hat\xi}'_t)}, 
\label{eq:4}
\end{eqnarray}
where $\delta T^2 = {\rm tr}({\hat T}{\hat\xi}_t{\hat T}{\hat\xi}_t)
-t^2$. As for the Fisher information, by use of the evolution equation 
${\hat\xi}'_t = -\ri({\hat H}{\hat\xi}_t-{\hat\xi}_t{\hat H})$, where we shall be 
working in units $\hbar=1$, we find 
\begin{eqnarray}
{\rm tr}({\hat\xi}'_t{\hat\xi}'_t) = 2 \left[ {\rm tr}({\hat H}^2 
{\hat\xi}^2_t) - {\rm tr}({\hat H}{\hat\xi}_t{\hat H}{\hat\xi}_t)
\right], \label{eq:5}
\end{eqnarray}
which is twice the skew information introduced by Wigner and Yanase 
\cite{WY}. 

For a pure state satisfying ${\hat\xi}_t^2={\hat\xi}_t$, the skew information 
is maximised and (\ref{eq:5}) reduces to twice the energy variance 
$2\Delta H^2$, whereas $\delta T^2=0$ for pure states. 
Thus for pure state we recover 
the standard uncertainty relation. For mixed states, however, we obtain a 
modified form of uncertainty relation 
\begin{eqnarray}
\Delta T^2 + \delta T^2 \geq \frac{1}{4(\Delta H^2 - \delta H^2)} . 
\label{eq:6}
\end{eqnarray}
Note that 
\begin{eqnarray}
\Delta H^2 - \delta H^2 = {\rm tr}({\hat H}^2{\hat\rho}) - {\rm tr}({\hat H}
\sqrt{\hat\rho} {\hat H}\sqrt{\hat\rho})  
\end{eqnarray}
appearing on the right side of (\ref{eq:6}) 
is the Wigner-Yanase skew information associated with ${\hat H}$, which is 
a positive quantity that is strictly smaller than the variance $\Delta H^2$ for 
mixed states (because $\Delta H^2 \geq \delta H^2 \geq 0$ for any state) and 
is equal to the variance for pure states (because $\delta H^2=0$ for pure 
states). The quantity appearing on the left side of (\ref{eq:6}), 
\begin{eqnarray}
\Delta T^2 + \delta T^2 = {\rm tr}({\hat T}^2{\hat\rho}) + {\rm tr}({\hat T}
\sqrt{\hat\rho} {\hat T}\sqrt{\hat\rho})  - 2\, \left({\rm tr}({\hat T}{\hat\rho})
\right)^2 , 
\label{eq:8}
\end{eqnarray}
is strictly greater than the variance $\Delta T^2$ for mixed states and is 
equal to the variance of ${\hat T}$ for pure states. We shall refer to 
(\ref{eq:8}) as the `skew information of the second kind'. 

What the generalised uncertainty relation (\ref{eq:6}) suggests is that for 
mixed states, perhaps the more appropriate measure of uncertainty error 
in parameter estimation is the skew information of the second kind, rather 
than the variance, whereas the sensitivity of the state in parameter variation, 
as measured by the Fisher information, is given by the Wigner-Yanase 
skew information, again rather than the variance. \\ 

\noindent \textbf{3. Geometric derivation of the quantum Cram\'er-Rao 
inequality} \\

The idea that we shall exploit in order to obtain higher-order corrections to 
the uncertainty relation (\ref{eq:6}) is as follows. We consider the Hilbert 
space ${\mathcal H}$ of square-root density matrices equipped with the 
trace inner product. Note that ${\mathcal H}$ is not the Hilbert space of 
states that describes the system; rather, it is a larger Hilbert space in which 
the density matrices are embedded via the square-root map. Thus, on 
${\mathcal H}$, the square-root of a mixed state ${\hat\rho}$ is interpreted 
like a `pure' state vector, where the inner product of two such vectors 
${\hat\xi}$ and ${\hat\eta}$ is defined by ${\rm tr}({\hat\xi}{\hat\eta})$. The 
advantage of working on ${\mathcal H}$ is that vectorial operations that 
have been used effectively in \cite{BH1,BH2,BH3} for statistical analysis of 
pure states can be extended into the domain of mixed states, albeit the 
calculations do become more elaborate. 

To begin, we shall consider level surfaces $t({\hat\xi})=c$ 
in ${\mathcal H}$ associated with 
constant expectation values of the operator ${\hat T}$: 
\begin{eqnarray}
t({\hat\xi})=\frac{{\rm tr}({\hat T}{\hat\xi}^2)}{{\rm tr}({\hat\xi}^2)} . 
\end{eqnarray}
Note that elements of ${\mathcal H}$ satisfy the condition 
${\rm tr}({\hat\xi}^2)<\infty$, and the normalisation condition 
${\rm tr}({\hat\xi}^2)=1$ is always imposed after performing calculations. 
In this way, projectively meaningful results can be obtained. Next, we 
consider the vector in ${\mathcal H}$ that is normal to the level surface 
$t({\hat\xi})=c$ at ${\hat\xi}$. A calculation shows that 
\begin{eqnarray}
{\hat\nabla} t=\frac{\partial t}{\partial {\hat\xi}}\bigg
\vert_{{\rm tr}({\hat\xi}^2)=1}
={\hat\xi}{\hat T}+{\hat T}\xi-2\, {\rm tr}({\hat T}{\hat\xi}^2)
\, {\hat\xi} , 
\end{eqnarray}
whose squared magnitude gives
\begin{eqnarray}
\big|{\hat\nabla} t\big|^{2}=2\left( {\rm tr}({\hat T}^{2}{\hat\xi}^2)+
{\rm tr}({\hat T}{\hat\xi}{\hat T}{\hat\xi})-2 \left( {\rm tr}({\hat T}{\hat\xi}^2)
\right)^2 \right)  . 
\label{eq:11}
\end{eqnarray}
Observe from (\ref{eq:8}) that this is precisely twice the left side of 
(\ref{eq:6}) when ${\hat\xi}$ is the solution to the dynamical equation 
${\hat\xi}'_t = -\ri({\hat H}{\hat\xi}_t-{\hat\xi}_t{\hat H})$. It follows that 
the uncertainties associated with estimating 
the parameter $t$ is given exactly by (rather than bounded by) half the 
squared magnitude of the normal vector ${\hat\nabla} t$ in ${\mathcal H}$. 

With this observation we are in the position to estimate the squared length 
of the vector ${\hat\nabla} t$. To this end we use the elementary fact that 
the squared norm of a vector is equal to the sum of the squared norms of 
its orthogonal components with respect to any choice of an orthonormal frame. 
To fix a frame for ${\mathcal H}$ we shall be using the vector ${\hat\xi}_t$ 
and its higher-order derivatives in $t$, and then orthogonalise them using 
the standard Gram-Schmidt scheme. The first step therefore is to identify 
the derivatives of the state. From ${\hat\xi}_{t}=\re^{-{\rm i}{\hat H} t}
{\hat\xi}_{0}\re^{{\rm i}{\hat H} t}$ we have 
${\hat\xi}_t'=\partial_{t}{\hat\xi}_t=\ri({\hat\xi}_t{\hat H}-{\hat H}{\hat\xi}_t)$. 
Another differentiation then gives 
\begin{eqnarray}
{\hat\xi}_t'' = -{\hat H}^{2}{\hat\xi}_t+2{\hat H}{\hat\xi}_t{\hat H}
-{\hat\xi}_t{\hat H}^{2}, 
\end{eqnarray}
and so on. More generally, a short calculation shows that 
\begin{eqnarray}
{\hat\xi}_t^{(n)}=(-\ri)^{n}\sum_{k=0}^{n} (-1)^k \binom{n}{k} 
{\hat H}^{n-k} \, {\hat\xi}_t \, {\hat H}^{k}. 
\label{eq:13}
\end{eqnarray}
The idea is to use these derivatives to form the proper velocity, acceleration, 
and higher-order analogues of them, which in turn form the basis for 
${\mathcal H}$. Letting $\{{\hat\Psi}_n\}$ denote the resulting orthogonal 
vectors, we thus find 
\begin{eqnarray}
\Delta T^2 + \delta T^2 = \frac{1}{2} \sum_{n=0}^\infty \frac{\left[{\rm tr}
\big(({\hat\xi}_t{\hat T}+{\hat T}{\hat\xi}_t-2t{\hat\xi}){\hat\Psi}_{n}\big)\right]^{2}}
{{\rm tr}({\hat\Psi}_{n}{\hat\Psi}_{n})} .
\label{eq:14}
\end{eqnarray}

The next step is to determine the set of vectors $\{{\hat\Psi}_n\}$. We let 
${\hat\Psi}_0={\hat\xi}_t$. We then see at once that ${\rm tr}
\big(({\hat\xi}_t{\hat T}+{\hat T}{\hat\xi}_t-2t{\hat\xi}){\hat\Psi}_0\big)=0$. In 
other words, ${\hat\Psi}_0={\hat\xi}_t$ is orthogonal to ${\hat\nabla}t$ so 
that the $n=0$ term in the sum (\ref{eq:14}) makes no contribution. For 
$n=1$ we let ${\hat\Psi}_1$ be the component of the first derivative 
${\hat\xi}_t'$ orthogonal to ${\hat\xi}_t$. However, because ${\rm tr}
({\hat\xi}'_t {\hat\xi}_t)=0$ we have ${\hat\Psi}_1={\hat\xi}_t'$ for the proper 
velocity vector. To calculate the $n=1$ contribution we need to calculate 
${\rm tr}\big(({\hat\xi}_t{\hat T}+{\hat T}{\hat\xi}_t){\hat\xi}_t'\big)$. In fact 
we have already deduced this trace (see just above equation (\ref{eq:4})), 
but let us calculate this explicitly. Using the dynamical equation 
${\hat\xi}'_t = -\ri({\hat H}{\hat\xi}_t-{\hat\xi}_t{\hat H})$ and the cyclic 
property of the trace operation we get 
${\rm tr}\big(({\hat\xi}_t{\hat T}+{\hat T}{\hat\xi}_t){\hat\xi}_t'\big)=
-{\rm tr}\big(\ri({\hat\xi}_t^2{\hat T}{\hat H}-{\hat\xi}_t^2{\hat H}{\hat T})\big)$, 
so from the commutation relation $\ri[{\hat H},{\hat T}]=1$ we see that 
the dependence on the estimator ${\hat T}$ drops out. Together with 
(\ref{eq:5}) we thus deduce (\ref{eq:6}) at once because the truncation of 
the sum at $n=1$ is necessarily smaller than or equal to the entire sum. 

This geometric derivation of (\ref{eq:6}) can be visualised as follows. In 
Hilbert space ${\mathcal H}$ we have a curve ${\hat\xi}_u$, where $u\geq0$. 
The arc length $s(t)$ of the curve is defined by 
\begin{eqnarray}
s(t) = \int_0^t |{\hat\xi}_u| \, \rd u = \sqrt{2(\Delta H^2 - \delta H^2)} \, t , 
\label{eq:15} 
\end{eqnarray}
which evidently is proportional to $t$ because the skew information of 
the energy is constant along the curve of unitary motion. 
Thus, the objective of time estimation is in effect to estimate 
the arc length of the curve. We now slice ${\mathcal H}$ by `space-like' 
hypersurfaces of constant expectations values of ${\hat T}$ along the 
curve. If the normal vector to the hypersurface at ${\hat\xi}_t$ is parallel 
to the tangent 
vector of the curve, then the estimation error is minimised and we have 
a minimum uncertainty state. Indeed, the unit normal vector to the 
surface at ${\hat\xi}_t$ is given by 
\begin{eqnarray}
\bm{\hat n} = \frac{{\hat\xi}_t{\hat T}+{\hat T}{\hat\xi}_t-2t{\hat\xi}}
{\sqrt{2(\Delta T^2+\delta T^2)}} , 
\label{eq:16} 
\end{eqnarray}   
whereas the unit tangent vector of the curve at ${\hat\xi}_t$ is given by 
\begin{eqnarray}
\bm{\hat e}_1 = \frac{ -\ri({\hat H}{\hat\xi}_t-{\hat\xi}_t{\hat H})}
{\sqrt{2(\Delta H^2-\delta H^2)}} . 
\label{eq:17} 
\end{eqnarray}   
The angular separation of these vectors, defined by $\cos\theta=
\bm{\hat n} \cdot \bm{\hat e}_1$, is thus given by 
\begin{eqnarray}
\theta = \cos^{-1} \left( \frac{1}{2\sqrt{(\Delta T^2+\delta T^2)
(\Delta H^2-\delta H^2)}} \right) . 
\end{eqnarray} 
For a minimum uncertainty state we have $\theta=0$, but in a generic 
case we have $\theta\neq0$. The angle $\theta$ thus can be viewed as the 
universal measure for the uncertainty associated with quantum state 
estimation in the case of a one-parameter family of quantum states. \\

\noindent \textbf{4. Quantum skew moments} \\

To work out the higher-order uncertainty bounds we shall encounter 
inner products of the form ${\rm tr}({\hat\xi}_t^{(m)}{\hat\xi}_t^{(n)})$, where 
$n+m$ is even. Before we proceed to examine the bounds more closely, 
let us analyse this inner product first 
because it leads to the notion of \textit{higher-order quantum skew 
central-moments}, first envisaged in \cite{DCB} but has not been worked 
out previously. Specifically, in our analysis it is the even-order skew 
moments that determine the higher-order corrections, in a way analogous 
to the bound given in (\ref{eq:2}). We shall define the $(n+m)$-th order 
skew central moment of the Hamiltonian by 
$S_{n+m}={\rm tr}({\hat\xi}_t^{(n)}{\hat\xi}_t^{(m)})$ if $n+m=4k+2$; 
whereas $S_{n+m}=-{\rm tr}({\hat\xi}_t^{(n)}{\hat\xi}_t^{(m)})$ if $n+m=4k$ for 
all $k=0,1,2,\ldots$. The intuition of the appearance of the minus sign here 
when $n+m$ is a multiple of four is as follows. For a central moment we 
demand that the highest-order moment ${\rm tr}({\hat H}^{n+m}{\hat\xi}^2)$ 
appearing inside to have the positive sign, but when the states are 
differentiated in the inner product ${\rm tr}({\hat\xi}_t^{(n)}{\hat\xi}_t^{(m)})$ 
we get a factor of $-\ri^{n+m}$ in front of the highest-order 
moment term. Hence we insert an extra minus sign to define the skew 
moments when $-\ri^{n+m}=-1$, or equivalently, when $n+m=4k$. 
With these preliminaries we have the following: 

\vspace{0.2cm} 
\noindent \textbf{Lemma}. The higher-order quantum skew central-moments 
are given by 
\begin{eqnarray}
S_{4L+2}=2 \sum_{k=0}^{2L}(-1)^{k}\binom{4L+2}{k}{\rm tr}
({\hat H}^{4L+2-k}{\hat\xi}_t{\hat H}^{k}{\hat\xi}_t) 
-\binom{4L+2}{2L+1}{\rm tr}({\hat H}^{2L+1}{\hat\xi}_t{\hat H}^{2L+1}{\hat\xi}_t)
\label{eq:23} 
\end{eqnarray}
and
\begin{eqnarray}
S_{4L}=2 \sum_{k=0}^{2L-1}(-1)^{k}\binom{4L}{k}{\rm tr}
({\hat H}^{4L-k}{\hat\xi}_t{\hat H}^{k}{\hat\xi}_t) +
\binom{4L}{2L}{\rm tr}({\hat H}^{2L}{\hat\xi}_t{\hat H}^{2L}{\hat\xi}_t)
\label{eq:24}
\end{eqnarray}
for all  $L>0$, whereas for $L=0$ we have $S_{0}=1$. 
\vspace{0.2cm} 

\noindent \textit{Proof}. 
To verify these, let us first consider the case $m+n=4L+2$. Then we have 
$n=4L+2-m$ and hence from (\ref{eq:13}) we find 
\begin{eqnarray} 
{\rm tr}({\hat\xi}_t^{(n)}{\hat\xi}_t^{(m)}) = - \sum_{k=0}^{4L+2-m} 
\sum_{l=0}^{m} (-1)^{k+l} \binom{4L+2-m}{k} \binom{m}{l} {\rm tr} \! 
\left( {\hat H}^{4L+2+l-k-m}{\hat\xi}_t{\hat H}^{m+k-l}{\hat\xi}_t \right) . 
\label{eq:25} 
\end{eqnarray} 
Now let $\alpha=m+k-l$. Then $\alpha$ ranges from 0 to $4L+2$ 
(specifically, $\alpha=0$ when $k=0$ and $l=m$; whereas $\alpha=4L+2$ 
when $l=0$ and $k=4L+2$). The trace term in (\ref{eq:25}) can then be written 
in the form ${\rm tr}({\hat H}^{4L+2-\alpha}{\hat\xi}_t{\hat H}^{\alpha}{\hat\xi}_t)$. 
For $\alpha=0$, corresponding to $(k,l)=(0,m)$, the the product of the 
binomial coefficients in (\ref{eq:25}) gives 1. For $\alpha=1$, corresponding 
to $(k,l)=(0,m-1)$ or $(k,l)=(1,m)$, the product of the binomial 
coefficients give $m$ or $4L+2-m$, both with the same sign, so they add up 
to give $4L+2=(4L+2)!/1!(4L+2-1)!$. For $\alpha=2$, corresponding 
to $(k,l)=(0,m-2)$, $(k,l)=(1,m-1)$, or $(k,l)=(2,m)$, the the product of 
the binomial coefficients give $m(m-1)$, $m(4L+2-m)$, or 
$(4L+2-m)(4L+2-m-1)/2$, again all with the same sign, so they add up 
to give $(4L+2)!/2!(4L+2-2)!$. Continuing along this line we observe that 
the double sum in (\ref{eq:25}) can be replaced with a single sum over 
$\alpha$, along with the binomial coefficient $(4L+2)!/\alpha!(4L+2-\alpha!)$ 
and with alternating signs: 
\begin{eqnarray} 
{\rm tr}({\hat\xi}_t^{(n)}{\hat\xi}_t^{(m)}) = (-1)^{m+1} \sum_{\alpha=0}^{4L+2} 
(-1)^{\alpha} \binom{4L+2}{\alpha} {\rm tr} \! 
\left( {\hat H}^{4L+2-\alpha}{\hat\xi}_t{\hat H}^{\alpha}{\hat\xi}_t \right) . 
\label{eq:26} 
\end{eqnarray} 
Notice that the sum contains an odd number of terms. Apart from the 
middle term in the sum for which $\alpha=2L+1$, each term appears 
twice on account of the symmetry of binomial coefficients, so we can 
truncate the sum in (\ref{eq:26}) at $\alpha=2L$, 
double the result, and add the term for $\alpha=2L+1$; this gives 
(\ref{eq:23}). An essentially identical line of argument then gives 
(\ref{eq:24}) when $m+n=4L$. 
\hfill $\Box$
\vspace{0.1cm} 

We note that the even-order skew moment $S_{2n}$ defined here does 
not reduce to the $n$-th central moment in the pure-state limit (except 
for $n=1$). To see this it suffices to consider $S_4$. Then from (\ref{eq:24}) 
we have
\begin{eqnarray}
S_{4}=2\left[ {\rm tr}({\hat H}^{4}{\hat\xi}_t^2) - 4 {\rm tr}({\hat H}^{3}{\hat\xi}_t
{\hat H}{\hat\xi}_t) + 3 {\rm tr}({\hat H}^{2}{\hat\xi}_t{\hat H}^{2}{\hat\xi}_t) 
\right] , 
\end{eqnarray}
and hence, for a pure state for which ${\hat\xi}={\hat\xi}^2$ we obtain 
\begin{eqnarray}
S_{4}=2\left[ \langle {\hat H}^{4}\rangle - 4 \langle{\hat H}^{3}\rangle 
\langle {\hat H}\rangle + 3 \langle{\hat H}^{2}\rangle^2 \right] .
\end{eqnarray}
Evidently, this is distinct from the fourth central moment of the 
Hamiltonian 
\begin{eqnarray}
\mu_{4}= \langle {\hat H}^{4}\rangle - 4 \langle{\hat H}^{3}\rangle 
\langle {\hat H}\rangle + 6 \langle{\hat H}^{2}\rangle\langle{\hat H}\rangle^2 
-3 \langle{\hat H}\rangle^4 . 
\end{eqnarray}
The fact that the skew moments do not reduce to central moments for 
pure states will have an implication when we compare the pure-state 
limit of the results obtained here to those obtained in \cite{BH3}. \\

\noindent \textbf{5. Higher-order variance bounds} \\

The decomposition (\ref{eq:14}) has one apparent disadvantage in that 
while the left side represents the magnitude of quadratic error resulting 
from using the estimator ${\hat T}$, the right side in general is also dependent 
on ${\hat T}$. For a statistically meaningful error bound we do not wish 
it to have a dependence on the choice of the estimator. On the other hand, 
the dependence on ${\hat T}$ dropped out in the contribution from 
the $n=1$ term, owing to the commutation relation of ${\hat T}$ and 
${\hat H}$. In general, we shall find that 
the dependence on ${\hat T}$ drops out from all the odd-order terms due 
to the commutation relation, whereas the even-order terms give rise to 
the anticommutator of ${\hat T}$ and ${\hat H}$, which we do not know 
how to evaluate. Thus the approach we have taken here gives rise to 
a generalisation of the so-called Robertson-Schr\"odinger uncertainty 
relation, where the uncertainty lower bound has terms proportional to the 
commutator, and terms proportional to the anticommutator. 
(An analogous structure emerges in the analysis of \cite{BH3} 
for pure states.) In what follows, therefore, we shall be focused on the 
odd-order terms that we can evaluate explicitly. 

The general Gram-Schmidt scheme for generating the orthogonal frame is 
given as follows: 
\begin{align}
{\hat\Psi}_{0}&={\hat\xi}_t \nonumber \\
{\hat\Psi}_{1}&={\hat\xi}_t'-\frac{{\rm tr}({\hat\xi}'{\hat\Psi}_{0})}
{{\rm tr}({\hat\Psi}_{0}{\hat\Psi}_{0})}{\hat\Psi}_{0} \nonumber \\
{\hat\Psi}_{2}&={\hat\xi}_t''-\frac{{\rm tr}({\hat\xi}_t''{\hat\Psi}_{1})}{{\rm tr}
({\hat\Psi}_{1}{\hat\Psi}_{1})}{\hat\Psi}_{1}-\frac{{\rm tr}({\hat\xi}_t''
{\hat\Psi}_{0})}{{\rm tr}({\hat\Psi}_{0}{\hat\Psi}_{0})}{\hat\Psi}_{0} \\
{\hat\Psi}_{3}&={\hat\xi}_t'''-\frac{{\rm tr}({\hat\xi}_t'''{\hat\Psi}_{2})}
{{\rm tr}({\hat\Psi}_{2}{\hat\Psi}_{2})}{\hat\Psi}_{2}-\frac{{\rm tr}({\hat\xi}_t^{'''}{\hat\Psi}_{1})}{{\rm tr}({\hat\Psi}_{1}{\hat\Psi}_{1})}{\hat\Psi}_{1}-
\frac{{\rm tr}({\hat\xi}_t'''{\hat\Psi}_{0})}{{\rm tr}({\hat\Psi}_{0}{\hat\Psi}_{0})}
{\hat\Psi}_{0} \nonumber  \\ &\,\,\,\vdots \nonumber 
\end{align}
We have already observed, however, that the inner product 
${\rm tr}({\hat\xi}'{\hat\Psi}_{0})$ appearing in ${\hat\Psi}_1$ vanishes so 
that ${\hat\Psi}_{1}={\hat\xi}_t'$. In 
fact, there are various other cancellations that simplify the expressions for 
the $\{{\hat\Psi}_n\}$. To this end let us first explore the inner product 
${\rm tr}({\hat\xi}_t^{(m)}{\hat\xi}_t^{(n)})$. This inner product vanishes when 
$n+m$ is an odd number. To see this, it suffices to consider the inner 
product  
${\rm tr}({\hat\xi}_t^{(2m)}{\hat\xi}_t^{(2n+1)})$ because the sum of two integers 
is odd only if one is even and the other is odd. Then from (\ref{eq:13}) we 
obtain 
\begin{eqnarray}
{\rm tr}\left({\hat\xi}_t^{(2m)}{\hat\xi}_t^{(2n+1)}\right) &=& -\ri (-1)^{m+n} 
\nonumber \\ && \hspace{-1.0cm}  \times 
\sum_{k=0}^{2m} \sum_{l=0}^{2n+1} (-1)^{k+l} \binom{2m}{k} \binom{2n+1}{l} 
{\rm tr}\left({\hat\xi} {\hat H}^{2n+1+k-l}{\hat\xi}{\hat H}^{2m-k+l}\right) . 
\label{eq:20} 
\end{eqnarray}
The vanishing of the double sum for all $m,n$ then follows from the symmetry 
of the binomial coefficients. Specifically, on inspection of the summand: 
\[ (-1)^{k+l} \frac{{\rm tr}\left({\hat\xi} {\hat H}^{2n+1+k-l}{\hat\xi}
{\hat H}^{2m-k+l}\right) }{k!\,(2m-k)!\, l!\, (2n+1-l)!} , \] 
where we have taken the overall multiplicable 
constant $(2m)!\,(2n+1)!$ outside of the sum, 
we observe the following. For each $(k,l)=(p,q)$ there is another term in 
the sum with $(k,l)=(2m-p,2n+1-q)$, which has an identical expression but 
with an opposite sign, when $p\neq m$. Thus they cancel. When $k=p=m$, 
there are even terms in the sum and for each term with $l=q$ there is another 
identical term with $l=2n+1-q$ having the opposite sign, so they also cancel. 
This shows that all the terms on the right side of (\ref{eq:20}) cancel, from 
which we deduce that ${\rm tr}({\hat\xi}_t^{(2m)}{\hat\xi}_t^{(2n+1)})=0$. The 
elements of the odd-order orthogonal frame that we are interested (the 
even terms give rise to anticommutator of ${\hat T}$ and ${\hat H}$) 
are therefore given by the set 
\begin{align}
{\hat\Psi}_{1}&={\hat\xi}_t^{(1)} \nonumber \\
{\hat\Psi}_{3}&={\hat\xi}_t^{(3)}-\frac{{\rm tr}({\hat\xi}_t^{(3)}{\hat\Psi}_{1})}
{{\rm tr}({\hat\Psi}_{1}{\hat\Psi}_{1})}{\hat\Psi}_{1} \nonumber \\
{\hat\Psi}_{5}&={\hat\xi}_t^{(5)}-\frac{{\rm tr}({\hat\xi}_t^{(5)}{\hat\Psi}_{3})}
{{\rm tr}({\hat\Psi}_{3}{\hat\Psi}_{3})}{\hat\Psi}_{3}-\frac{{\rm tr}({\hat\xi}_t^{(5)}{\hat\Psi}_{1})}{{\rm tr}({\hat\Psi}_{1}{\hat\Psi}_{1})}{\hat\Psi}_{1}   \label{eq:21}  \\
{\hat\Psi}_{7}&={\hat\xi}_t^{(7)}-\frac{{\rm tr}({\hat\xi}_t^{(7)}{\hat\Psi}_{5})}
{{\rm tr}({\hat\Psi}_{5}{\hat\Psi}_{5})}{\hat\Psi}_{5}-\frac{{\rm tr}({\hat\xi}_t^{(7)}
{\hat\Psi}_{3})}{{\rm tr}({\hat\Psi}_{3}{\hat\Psi}_{3})}{\hat\Psi}_{3}-\frac{{\rm tr}
({\hat\xi}_t^{(7)}{\hat\Psi}_{1})}{{\rm tr}({\hat\Psi}_{1}{\hat\Psi}_{1})}{\hat\Psi}_{1} 
\nonumber \\ &\,\,\,\vdots \nonumber 
\end{align}

Our next step is to work out the norms $N_n={\rm tr}({\hat\Psi}_n{\hat\Psi}_n)$ 
appearing in (\ref{eq:21}). For $n=1$ this is given in (\ref{eq:5}) by twice 
the Wigner-Yanase skew information of the energy. Let us examine 
the case $n=3$. From the expression for ${\hat\Psi}_3$ in (\ref{eq:21}) and 
the orthogonality of the vectors $\{{\hat\Psi}_n\}$ we see at once that 
\begin{eqnarray}
{\rm tr}({\hat\Psi}_{3}{\hat\Psi}_{3})={\rm tr}({\hat\xi}_t^{(3)}{\hat\xi}_t^{(3)})
-\frac{[{\rm tr}({\hat\xi}_t^{(3)}{\hat\xi}_t^{(1)})]^{2}}
{{\rm tr}({\hat\Psi}_{1}{\hat\Psi}_{1})}.
\end{eqnarray}
More generally, on a closer inspection of (\ref{eq:21}) we deduce that the 
norm $N_n$ can be constructed recursively from the norms of the 
lower-order vectors along with the inner products of the form 
${\rm tr}({\hat\xi}_t^{(m)}{\hat\xi}_t^{(n)})$ where $n+m$ is even. These inner 
products, however, are given by the higher-order skew moments. 

With these observations at hand, 
let us now work out the expressions for the norms $N_n=
{\rm tr}({\hat\Psi}_n{\hat\Psi}_n)$, $n=1,3,5,\ldots$, in terms of the skew 
moments. In fact, a short calculation at once reveals the recursive 
structure of these norms, for, we have 
\begin{eqnarray} 
N_1 = S_2, \qquad N_3 = \frac{\begin{vmatrix} S_6 & S_4 \\ 
S_4 & S_2 \end{vmatrix}}{S_2}, \qquad N_5 = \frac{\begin{vmatrix} 
S_{10} & S_8 & S_6 \\ S_8 & S_6 & S_4 \\ 
S_6 & S_4 & S_2 \end{vmatrix}}{\begin{vmatrix} S_6 & S_4 \\ 
S_4 & S_2 \end{vmatrix}}, 
\end{eqnarray} 
and so on. The norms are therefore given by the ratios of the determinants 
of the matrices of even-order central moments. Let us denote these 
determinants by 
$D_{2n}$: 
\begin{eqnarray} 
D_{2n}=
\begin{vmatrix}
S_{2n} & S_{2n-2} & \dots & S_{n+1} \\
S_{2n-2} & S_{2n-4} & \dots & S_{n-1} \\
\vdots & \vdots & \ddots & \vdots \\
S_{n+1} & S_{n-1} & \dots & S_{2} \\
\end{vmatrix} 
\label{eq:29} 
\end{eqnarray} 
for $n=1,3,5,\ldots$. Then with the convention that $D_{-2}=1$, we 
have 
\begin{eqnarray} 
N_{n}=\frac{D_{2n}}{D_{2n-4}}  
\label{eq:32} 
\end{eqnarray} 
for these odd-order norms. 
The overall structure of the normalisation $\{N_n\}$ thus is identical to the 
one obtained in \cite{BH3} for pure states, except that the norms obtained 
here are different from the ones obtained in \cite{BH3} because the skew 
moment $S_n$ defined here does not reduce to the $n$-th central moment 
$\mu_n$ in the pure-state limit, except for $n=2$, as indicated above. 

The structural similarity of the determinant in (\ref{eq:29}) to the one 
obtained in \cite{BH3} involving central moments, however, implies that 
the geometric interpretation discussed in \cite{BH3} is applicable in the 
present case. Namely, regarding the state ${\hat\xi}_t$ as a curve in 
Hilbert space, $D_{2n}$ for each $n$ can 
be expressed as the trace norm of the tensor obtained by a totally 
skew-symmetric product of the vectors ${\hat\xi}_t^{(1)}$, ${\hat\xi}_t^{(3)}$, 
$\cdots$, ${\hat\xi}_t^{(n-2)}$, ${\hat\xi}_t^{(n)}$. It follows that the norms 
$\{N_n\}$ can be interpreted as representing higher-order torsions of the 
curve in Hilbert space. 

We recall that our objective is to determine the right side of (\ref{eq:14}) 
over odd $n$. Having identified the denominators in (\ref{eq:14}), let us 
turn our attention to the numerators. On account of the fact that ${\hat\xi}_t$ 
and ${\hat\Psi}_n$ are orthogonal, the term containing the parameter $t$ 
drops out and the numerator reduces to the square of 
${\rm tr}\big(({\hat\xi}_t{\hat T}+{\hat T}{\hat\xi}_t){\hat\Psi}_{n}\big)$; but 
${\hat\Psi}_{n}$ in turn is expressed as a linear combination of odd-order 
derivatives of ${\hat\xi}$. Before we proceed further, let us first establish 
the following: 

\vspace{0.2cm} 
\noindent \textbf{Lemma}. The inner product 
${\rm tr}\big(({\hat\xi}_t{\hat T}+{\hat T}{\hat\xi}_t){\hat\Psi}_{n}\big)$ is 
independent of ${\hat T}$ for $n$ odd, whereas for $n$ even it is 
determined by the anticommutators of ${\hat T}$ and ${\hat H}^p$ for 
a range of $p$. 
\vspace{0.2cm} 

\noindent \textit{Proof}. 
Let us first consider the case where $n$ is odd. Then ${\hat\Psi}_{n}$ is 
expressed as a linear combination of odd-order derivatives of the state 
${\hat\xi}_t$, so let us examine more explicitly an expression of the form 
\begin{eqnarray} 
{\rm tr}\big({\hat\xi}_t^{(2m+1)}({\hat\xi}_t{\hat T}+{\hat T}{\hat\xi}_t)\big) = 
(-1)^{n+1} \ri \sum_{k=0}^{2m+1} (-1)^k \binom{2m+1}{k} {\rm tr}\big( 
{\hat H}^{2m+1-k}{\hat\xi}_t{\hat H}^k(
{\hat\xi}_t{\hat T}+{\hat T}{\hat\xi}_t)\big) ,
\end{eqnarray} 
where we have made use of (\ref{eq:13}). On a closer inspection we find 
that for $k=\alpha$ the right side is 
\[ 
(-1)^{m+1} \ri (-1)^\alpha \frac{(2m+1)!}{\alpha!(2n+1-\alpha)!} 
{\rm tr}\big( {\hat T}{\hat H}^{2m+1-\alpha}{\hat\xi}_t{\hat H}^\alpha
{\hat\xi}_t + {\hat H}^\alpha {\hat T}{\hat\xi}_t {\hat H}^{2m+1-\alpha} 
{\hat\xi}_t\big) ,
\] 
whereas for $k=2m+1-\alpha$ it gives 
\[ 
-(-1)^{m+1} \ri (-1)^\alpha \frac{(2m+1)!}{\alpha!(2n+1-\alpha)!} 
{\rm tr}\big( {\hat H}^{2m+1-\alpha}{\hat T}{\hat\xi}_t{\hat H}^\alpha
{\hat\xi}_t + {\hat T}{\hat H}^\alpha {\hat\xi}_t {\hat H}^{2m+1-\alpha} 
{\hat\xi}_t\big) .
\] 
Adding these together we obtain 
\[ 
(-1)^{m+1-\alpha} \frac{(2m+1)!}{\alpha!(2n+1-\alpha)!} \left[ 
{\rm tr}\big( \ri[{\hat T},{\hat H}^{2m+1-\alpha}] {\hat\xi}_t{\hat H}^\alpha 
{\hat\xi}_t\big) + {\rm tr}\big( \ri[{\hat H}^{\alpha},{\hat T}] 
 {\hat\xi}_t{\hat H}^{2m+1-\alpha} {\hat\xi}_t\big) \right] .
\] 
We therefore see that the dependence on the estimator ${\hat T}$ 
drops out on account of the commutation relation $\ri[{\hat H}^p,{\hat T}]
= p{\hat H}^{p-1}$, and we obtain 
\begin{eqnarray} 
{\rm tr}\big({\hat\xi}_t^{(2m+1)}({\hat\xi}_t{\hat T}+{\hat T}{\hat\xi}_t)\big) &=& 
\sum_{\alpha=0}^m (-1)^{m+1-\alpha} \frac{(2m+1)!}{\alpha!(2m+1-\alpha)!} 
\nonumber \\ && \hspace{-1.20cm} \times 
\left[ \alpha \, {\rm tr}({\hat H}^{\alpha-1}{\hat\xi}_t{\hat H}^{2m+1-\alpha}
{\hat\xi}_t) - (2m+1-\alpha) {\rm tr}({\hat H}^{2m-\alpha}{\hat\xi}_t
{\hat H}^{\alpha}{\hat\xi}_t) \right] .   
\label{eq:34} 
\end{eqnarray} 
This establishes the claim that the dependence of the bound on the 
estimator ${\hat T}$ drops out from the odd-order terms on account of the 
commutation relation. When $n$ is even, however, this argument shows 
that the two corresponding terms for $k=\alpha$ and $k=2m-\alpha$ add 
up to give anticommutators of the form $\{{\hat H}^p,{\hat T}\}$; while for 
$k=m$ the two terms resulting from ${\hat\xi}_t{\hat T}+{\hat T}{\hat\xi}_t$ 
also gives an anticommutator of ${\hat T}$ and ${\hat H}^m$. 
\hfill $\Box$
\vspace{0.1cm} 

On a closer examination we see that the expression in the right side of 
(\ref{eq:34}) can in fact be simplified further.  To see this we note that the 
first sum in (\ref{eq:34}) can be expressed alternatively in the form
\[
\sum_{\alpha=0}^{m-1} (-1)^{m-\alpha}
\frac{(2m+1)!}{\alpha!(2m-\alpha)!} 
\, {\rm tr}({\hat H}^{\alpha}{\hat\xi}_t{\hat H}^{2m-\alpha}
{\hat\xi}_t) 
\]
by shifting the summation variable $\alpha\to\alpha+1$. On the other, the 
second sum in (\ref{eq:34}) can be expressed alternatively in the form 
\[ 
\sum_{\alpha=0}^{m-1} (-1)^{m-\alpha} \frac{(2m+1)!}{\alpha!(2m-\alpha)!} 
\, {\rm tr}({\hat H}^{2m-\alpha}{\hat\xi}_t {\hat H}^{\alpha}{\hat\xi}_t) - 
\frac{(2m+1)!}{(m!)^2} \, 
{\rm tr}({\hat H}^{m}{\hat\xi}_t {\hat H}^{m}{\hat\xi}_t) . 
\]
Therefore, the two sums combine, and if we compare the result thus 
obtained with the expressions in (\ref{eq:23}) and (\ref{eq:24}) we 
deduce that 
\begin{eqnarray} 
{\rm tr}\big({\hat\xi}_t^{(2m+1)}({\hat\xi}_t{\hat T}+{\hat T}{\hat\xi}_t)\big) = 
(-1)^{m+2} \, (2m+1)\, S_{2m} .   
\label{eq:35} 
\end{eqnarray} 

With these results at hand we are now in the position to examine the 
numerator terms in (\ref{eq:14}). Our approach will be to deduce a recursion 
relation for the numerator $U_{n}:={\rm tr}(({\hat\xi}{\hat T}+{\hat T}{\hat\xi})
{\hat\Psi}_{n})$, because ${\hat\Psi}_{n}$ is expressed in terms of a linear 
combination of ${\hat\xi}_t^{(n)}$ and ${\hat\Psi}_{k}$ with $k=n-2,n-4,\ldots$. 
To deduce a recursion relation we need to work out the coefficients 
\begin{eqnarray}
F_{n,k} := \frac{{\rm tr}({\hat\xi}_t^{(n)}{\hat\Psi}_{k})}{{\rm tr}({\hat\Psi}_{k}
{\hat\Psi}_{k})} \, , \qquad (k=1,3,5,\ldots,n-2) 
\end{eqnarray} 
of ${\hat\Psi}_{k}$ in ${\hat\Psi}_{n}$. In terms of these coefficients we 
therefore have 
\begin{eqnarray}
{\hat\Psi}_{n}={\hat\xi}_t^{(n)}-\sum_{k=1,3,5,...}^{n-2}F_{n,k}{\hat\Psi}_{k} . 
\end{eqnarray}
A calculation analogous to the ones 
outlined above then shows that 
\begin{eqnarray}
F_{n,k}=\frac{(-1)^{\frac{1}{2}(n+k)-1}}{D_{2k}}
\begin{vmatrix}
S_{n+k} & S_{n+k-2} & \dots & S_{n+1} \\
S_{2k-2} & S_{2k-4} & \dots & S_{k-1} \\
\vdots & \vdots & \ddots & \vdots \\
S_{k+1} & S_{k-1} & \dots & S_{2} \\
\end{vmatrix} . 
\end{eqnarray}
Therefore, it follows at once that the recursion relation satisfied by the 
numerator terms is given by 
\begin{eqnarray}
U_{n}=(-1)^{\frac{1}{2}(n-1)}nS_{n-1}-\sum_{k=1,3,5,...}^{n-2}F_{n,k}U_{k},  \label{eq:39}
\end{eqnarray}
along with the initial condition $U_1=1$. 

Putting these together, we finally deduce that the generalised uncertainty 
relation takes the form 
\begin{eqnarray}
\Delta {\hat T}^{2}+\delta {\hat T}^{2} \geq 
\frac{1}{2} \sum_{n=1,3,5,...}\frac{U_{n}^2}{N_{n}} , 
\end{eqnarray} 
where the denominator $N_n$ is given by the ratio of matrix determinant 
(\ref{eq:32}) and the numerator $U_n$ is obtained recursively by (\ref{eq:39}). 
In this way, we are able to work out as many higher-order corrections 
as we wish to the generalised uncertainty relation associated with generic 
mixed-state density matrices. For example, truncating the sum at $n=3$ we 
deduce that 
\begin{equation}
(\Delta {\hat T}^{2}+\delta {\hat T}^{2})(\Delta {\hat H}^{2}-\delta {\hat H}^{2}) 
\geq \frac{1}{4} \left( 1 + \frac{(S_4-3S_2^2)^2}{S_6 S_2 - S_4^2} \right) .
\label{eq:41} 
\end{equation}
The structure of the bound here is therefore identical to that of (\ref{eq:2}), 
except that in the pure-state limit (\ref{eq:41}) does not reduce to (\ref{eq:2}), 
because while the left side becomes $\Delta {\hat T}^2 \Delta {\hat H}^2$, 
the higher-order skew moments $S_4$ and $S_6$ appearing on the right 
side of (\ref{eq:41}) do not reduce to central moments. 

In summary, we have worked out the higher-order corrections, in the sense 
of orders of the moments, to the uncertainty lower bound associated with a 
generic mixed state density matrix using the Hilbert space method. We have 
shown that an appropriate measure of 
estimation uncertainty, in the case of mixed quantum states, should be the 
skew information of the second kind introduced here, as opposed to the 
conventional variance measure. In contrast, the 
uncertainty lower bounds are expressed in terms of the skew moments of 
the first kind, which we have worked out explicitly. The notion of these skew 
moments of both kinds arise naturally in analysing mixed quantum states, 
and has no analogue when dealing with pure states.

\vspace{0.2cm}
\begin{footnotesize}
\noindent {\bf Acknowledgements}. DCB acknowledges support from the 
Russian Science Foundation, grant 20-11-20226.  The authors thank 
D.~O.~Soares-Pinto for drawing our attention to \cite{Diogo} in which 
the problem addressed here has been investigated and 
closely-related results have been obtained independently. 
\end{footnotesize}

\end{document}